\documentclass[aps,prd,twocolumn,nofootinbib,superscriptaddress,floatfix,notitlepage]{revtex4-1}

\usepackage{graphicx}
\usepackage{amsmath,amsfonts,amssymb}
\usepackage{hyperref}
\usepackage{color}
\usepackage{verbatim}

\newcommand{\td}{{\rm d}}

\begin{document}

\title{Gravitational waves from colliding vacuum bubbles in gauge theories}

\author{Marek Lewicki}
\email{marek.lewicki@fuw.edu.pl}
\affiliation{Faculty of Physics, University of Warsaw ul.\ Pasteura 5, 02-093 Warsaw, Poland}
\author{Ville Vaskonen}
\email{vvaskonen@ifae.es}
\affiliation{Institut de Fisica d'Altes Energies (IFAE), The Barcelona Institute of Science and Technology, Campus UAB, 08193 Bellaterra (Barcelona), Spain}

\begin{abstract}
We study production of gravitational waves (GWs) in strongly supercooled cosmological phase transitions in gauge theories. We extract from two-bubble lattice simulations the scaling of the GW source, and use it in many-bubble simulations in the thin-wall limit to estimate the resulting GW spectrum. We find that in presence of the gauge field the GW source decays with bubble radius as $\propto R^{-3}$ after collisions. This leads to a GW spectrum that follows $\Omega_{\rm GW} \propto \omega^{2.3}$ at low frequencies and $\Omega_{\rm GW} \propto \omega^{-2.4}$ at high frequencies, marking a significant deviation from the popular envelope approximation.
\end{abstract}

\maketitle

\section{Introduction}
We are currently witnessing the dawn of a new era in astrophysics and cosmology, started by the LIGO/Virgo observations of gravitational waves (GWs) from black hole  mergers~\cite{TheLIGOScientific:2017qsa,Abbott:2020niy}. Many experiments are planned to further explore GWs in a broad frequency range in the coming decades~\cite{Punturo:2010zz,Hild:2010id,Janssen:2014dka,Graham:2016plp,Audley:2017drz,Graham:2017pmn,Badurina:2019hst,Bertoldi:2019tck}. In addition to transient GW signals, such as those from black hole mergers, these experiments are able to probe stochastic GW backgrounds. In fact, recent results from NANOGrav pulsar timing observations~\cite{Arzoumanian:2020vkk} may already indicate the first observation of a stochastic GW background~\cite{Ellis:2020ena,Blasi:2020mfx,Vaskonen:2020lbd,DeLuca:2020agl,Nakai:2020oit,Ratzinger:2020koh,Kohri:2020qqd,Vagnozzi:2020gtf,Neronov:2020qrl,Middleton:2020asl}.

Observations of stochastic GW backgrounds could allow us a glimpse of the very early Universe as many high-energy processes are predicted to be potential sources of such backgrounds. In this paper we will focus on cosmological first-order phase transitions, which are one example of such a source~\cite{Witten:1984rs}. Many beyond Standard Model scenarios predict first-order phase transitions and a significant amount of work has already been put into the possibility of exploring them through GWs~\cite{Grojean:2006bp,Espinosa:2008kw,Dorsch:2014qpa,Jaeckel:2016jlh,Jinno:2016knw,Chala:2016ykx,Chala:2018opy,Artymowski:2016tme,Hashino:2016xoj,Vaskonen:2016yiu,Dorsch:2016nrg,Beniwal:2017eik,Baldes:2017rcu,Marzola:2017jzl,Kang:2017mkl,Iso:2017uuu,Chala:2018ari,Bruggisser:2018mrt,Megias:2018sxv,Croon:2018erz,Alves:2018jsw,Baratella:2018pxi,Angelescu:2018dkk,Croon:2018kqn,Brdar:2018num,Beniwal:2018hyi,Breitbach:2018ddu,Marzo:2018nov,Baldes:2018emh,Prokopec:2018tnq,Fairbairn:2019xog,Helmboldt:2019pan,Dev:2019njv,Ellis:2019flb,Jinno:2019bxw,Ellis:2019tjf,Azatov:2019png,vonHarling:2019gme,DelleRose:2019pgi,Mancha:2020fzw,Vanvlasselaer:2020niz,Giese:2020znk,Hoeche:2020rsg,Baldes:2020kam,Croon:2020cgk,Ares:2020lbt,Cai:2020djd,Bigazzi:2020avc,Wang:2020zlf}. 

In a first-order phase transition the Universe starts in a metastable false vacuum. The transition proceeds via nucleation and subsequent expansion of bubbles of the true vacuum~\cite{Coleman:1977py,Callan:1977pt,Linde:1981zj}. Eventually these bubbles collide and convert the whole Hubble volume into the new phase. In this process GWs are sourced by the bubble collisions~\cite{Kosowsky:1992vn,Cutting:2018tjt,Ellis:2019oqb,Lewicki:2019gmv,Cutting:2020nla,Lewicki:2020jiv} and plasma motions generated by the interactions of the plasma with the bubble walls~\cite{Kamionkowski:1993fg,Hindmarsh:2015qta,Hindmarsh:2016lnk,Hindmarsh:2017gnf,Ellis:2018mja,Hindmarsh:2019phv,Ellis:2020awk}. In strongly supercooled transitions the former source dominates~\cite{Ellis:2019oqb,Ellis:2020nnr}.

For the calculation of the GWs from colliding vacuum bubbles the equations of motion of the fields sourcing GWs need to be solved, requiring, in principle, 3D lattice simulations~\cite{Child:2012qg,Cutting:2018tjt,Cutting:2020nla}. These simulations are computationally very expensive as very large simulation volumes are needed in order to simulate multiple bubbles, and very dense lattices to resolve the thinning bubble walls. Therefore, it is practical to develop approximations that provide a realistic description of the phase transition dynamics and an accurate estimate of the resulting GW spectrum, but are computationally less expensive than full 3D lattice simulations.

For a long time the envelope approximation, introduced in Ref.~\cite{Kosowsky:1992vn} and studied further in Refs.~\cite{Huber:2008hg,Weir:2016tov,Jinno:2016vai}, has been used to estimate the GW spectrum sourced by the bubble collisions.  In this approximation the collided parts of the bubble walls are completely neglected and the GW spectrum is calculated in the thin-wall limit. Improved modeling was developed in Refs.~\cite{Jinno:2017fby,Konstandin:2017sat,Jinno:2019jhi,Jinno:2020eqg} as an attempt to model the behaviour of the plasma after the transition. Following a similar approach in Ref.~\cite{Lewicki:2020jiv} we developed a new estimate for the GW spectrum from bubble collisions by accounting for the scaling of the GW source after the collisions. Our estimate lead to a spectrum significantly different from the envelope approximation.

In this paper we consider a class of realistic models where bubble collisions can give the dominant contribution to the GW production. Furthermore, we describe breaking of a gauge U(1) symmetry, and study with lattice simulations the evolution of the scalar and gauge fields in two-bubble collisions. We find that the gradients in the complex phase of the scalar field are quickly damped after the collision by the gauge field. As a result, in gauge theories the GW source after the collision scales similarly to the case of just a real scalar, and the resulting GW spectrum follows $\propto\omega^{2.3}$ at low frequencies with a $\propto\omega^{-2.4}$ fall above the peak.

\begin{figure*}
\centering
\includegraphics[width=0.88\textwidth]{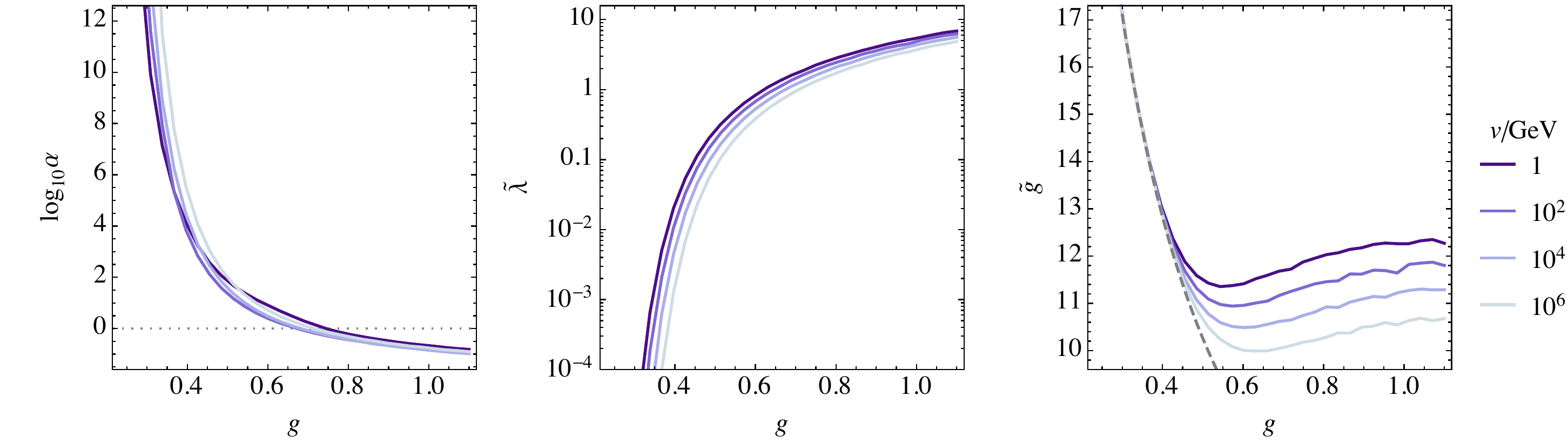}
\caption{The strength of the transition $\alpha$, and the dimensionless parameters $\tilde{\lambda}$ and $\tilde{g}$ (see Eq.~\eqref{eq:kappaG}) as a function of the gauge coupling $g$. Different curves correspond to different values of $v$. The dashed curve in the right panel shows $T=0$ limit of $\tilde{g}$.}
\label{fig:P}
\end{figure*}

\section{Phase transition} 
In order for the bubble collisions to give the dominant GW source, the phase transition has to be strongly supercooled~\cite{Ellis:2019oqb,Ellis:2020nnr}. Such strong supercooling is not typically realized in models with a polynomial scalar potential~\cite{Ellis:2018mja,Ellis:2019oqb}. Instead, in models featuring classical scale invariance~\cite{Randall:2006py,Konstandin:2011dr,Konstandin:2011ds,Jinno:2016knw,Iso:2017uuu,vonHarling:2017yew,Kobakhidze:2017mru,Marzola:2017jzl,Prokopec:2018tnq,Hambye:2018qjv,Marzo:2018nov,Baratella:2018pxi,Bruggisser:2018mrt,vonHarling:2019gme,Aoki:2019mlt,DelleRose:2019pgi,Fujikura:2019oyi,Wang:2020jrd} the transition can be so strongly supercooled that the interactions of the bubble wall with the plasma can be neglected~\cite{Ellis:2019oqb,Ellis:2020nnr}. Many such models also include a gauge U(1) symmetry under which the scalar field is charged, and the dominant contribution on the effective potential arises from the gauge field loops. The phase transition in these models is therefore similar to that in classically conformal scalar electrodynamics, which we choose as a benchmark model.

Scalar electrodynamics is described by the gauge U(1) symmetric Lagrangian
\begin{equation}
\mathcal{L} = -\frac14 (F_{\mu\nu})^2 + |D_\mu \phi|^2 - V(|\phi|) \,,
\end{equation}
where $F_{\mu\nu} = \partial_\mu A_\nu - \partial_\nu A_\mu$ and $D_\mu = \partial_\mu + ig A_\mu$ are the electromagnetic field strength tensor and the gauge covariant derivative. In classically conformal models the tree-level scalar potential is quartic, $V(|\phi|) = \lambda |\phi|/4$. A non-trivial minimum is revealed when the radiative corrections are taken into account~\cite{Coleman:1973jx}, and finite temperature effects induce a potential energy barrier between the symmetric and the symmetry-breaking minima. The one-loop effective potential of classically conformal scalar electrodynamics is
\begin{equation} \label{eq:pot0}
V(|\phi|) = \frac{g^2}{2} T^2 |\phi|^2 + \frac{3 g^4}{4\pi^2} |\phi|^4 \left[\ln\frac{|\phi|^2}{v^2} - \frac{1}{2} \right] \,,
\end{equation}
where $T$ denotes temperature of the plasma and $v$ the vacuum expectation value of $|\phi|$ at $T=0$.

The symmetric and broken vacua are degenerate at a critical temperature $T = T_c$. The bubble nucleation temperature $T_n < T_c$ is defined as the temperature at which the probability of nucleating at least one bubble in a horizon volume in a Hubble time approaches unity~\cite{Linde:1981zj}. In the left of Fig.~\ref{fig:P} we show the parameter $\alpha \equiv \Delta V(T=0)/\rho_{\rm rad}(T)$, that characterizes the strength of the transition, as a function of $g$ for different values of $v$. We assume that only vacuum and radiation energy densities, $\Delta V(T=0)$ and $\rho_{\rm rad}(T)$, contribute to the expansion rate, and approximate the effective number of relativistic degrees of freedom by its Standard Model value~\cite{Saikawa:2018rcs}. If $\alpha>1$ the transition finishes only after a vacuum energy dominated period. By strong supercooling we refer to $\alpha\gg 1$.

For the following analysis we define dimensionless parameters $\tilde{g}$ and $\tilde{\lambda}$ as 
\begin{equation} \label{eq:kappaG}
\tilde{g} = \frac{g v^2}{\sqrt{\Delta V}} \,, \qquad \tilde{\lambda} = \frac{g^2 v^2 T^2}{2 \Delta V} \,,
\end{equation}
such that $\tilde{\lambda}$ determines the shape of the scalar potential and $\tilde{g}$ the strength of the coupling between the gauge field and the scalar field. In the middle and right panels of Fig.~\ref{fig:P} we show these parameters at $T=T_n$. For strongly supercooled transitions $\tilde{g}^2 \approx 8\pi/(3 g^2)$ and $\tilde{\lambda}^2 \approx 60/(g_*(T_n)\alpha)$.

\begin{figure*}
\centering
\includegraphics[width=0.96\textwidth]{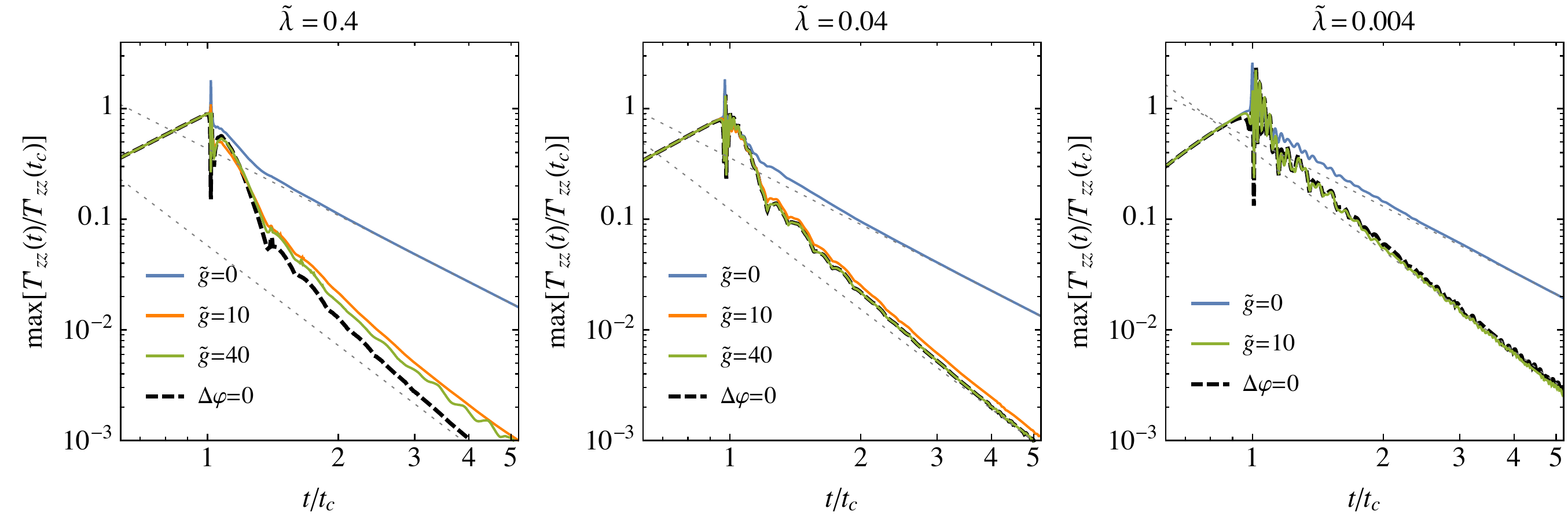}
\caption{Evolution of the GW source in collision of two bubbles averaged over simulations with different initial complex phase differences. The collision happens at $t=t_c$. Solid curves correspond to different values of $\tilde{g}$ and the dashed curve to $\Delta\varphi=0$. The dotted lines show $\propto t^{-2}$ and $\propto t^{-3}$ power-laws.}
\label{fig:scaling}
\end{figure*}

\section{Gravitational wave source}
Next we study two-bubble collisions in order to find how the GW source scales after the collision. The total energy spectrum in a direction $\hat{k}$ at an angular frequency $\omega = |\vec{k}|$ of the GWs emitted in the phase transition is given by~\cite{Weinberg:1972kfs}
\begin{equation} \label{eq:dEkdw}
\frac{\td E}{\td\Omega_k\td \omega} = 2G\omega^2 \Lambda_{ij,lm}(\hat{k}) T_{ij}^*(\vec{k}) T_{lm}(\vec{k}) \,,
\end{equation}
where $\Lambda_{ij,lm}$ is the transverse-traceless projection tensor. As $\Lambda_{ij,lm}\delta_{ij} = 0$, the part of the stress energy tensor that is proportional to the metric tensor $g_{\mu\nu}$ does not contribute to formation of GWs. We therefore define $T_{\mu\nu}$ as (see Appendix~\ref{lattice} for an explicit form)
\begin{equation}
T_{\mu\nu} \equiv \left(\frac{\partial \mathcal{L}}{\partial(\partial^\mu \phi)} \partial_\nu \phi + {\rm c.c}\right)  + \frac{\partial \mathcal{L}}{\partial(\partial^\mu A_\alpha)} \partial_\nu A_\alpha \,.
\end{equation}
The evolution of the system is governed by the equations of motion, given in the Lorentz gauge ($\partial_\mu A^\mu = 0$)\footnote{Our results are independent of the gauge choice because $T_{\mu\nu}$ is gauge invariant.} by
\begin{equation}\begin{aligned} \label{eq:EOM0}
&\Box A_\mu = i g (\phi^* \partial_\mu \phi - \phi \partial_\mu \phi^*) - 2 g^2 A_\mu |\phi|^2 \,, \\
&\Box \phi + \frac{\td V}{\td \phi^*} = -i2g A_\mu \partial^\mu \phi + g^2 A^2 \phi \,, 
\end{aligned}\end{equation}
which we solve on a lattice starting from a configuration where $A_\mu = 0$~\footnote{The thermal mass $\propto g^2 T^2 A^2$ stabilises the initial configuration without significantly affecting subsequent dynamics.} and two O(4) symmetric scalar field bubbles\footnote{The late evolution of the bubbles does not depend on whether the initial bubbles are O$(3)$ or O(4) symmetric~\cite{Lewicki:2019gmv}.} have nucleated simultaneously with their centers lying on $z$-axis~(see Appendix~\ref{lattice} for details of the lattice simulation). Then, along the collision axis only the $zz$ component of $T_{ij}$ is non-zero,
\begin{equation}
\begin{aligned}
T_{zz} =& \,2|\partial_z \phi| - (\partial_z A_t - \partial_t A_z) \partial_z A_t \\ &- i g A_z (\phi^*\partial_z\phi - \phi\partial_z\phi^*) \,.
\end{aligned}
\end{equation}

The bubble nucleation breaks the U(1) symmetry inside the bubble, as the complex phase of the scalar field, which we denote by $\varphi$ (i.e. $\phi = |\phi| e^{i\varphi}$), takes a value in the range $0\leq \varphi < 2\pi$.\footnote{As the gradients in the complex phase would increase the energy of the bubble, in the lowest energy configuration, and therefore for the nucleating bubbles, $\varphi$ is constant.} Eventually, as the bubbles expand, they will collide with bubbles containing different complex phases. Therefore, to get the average scaling of the GW source, we average $T_{zz}$ over simulations with different initial complex phase differences.

Our lattice simulations show that a $T_{zz}$ deviates from zero in a very narrow region around the bubble wall and this feature continues propagating almost at the speed of light after the collision. In Fig.~\ref{fig:scaling} we show by the solid curves the scaling of the maximal $T_{zz}$ as a function of time, which much after nucleation is obtained roughly at $z = \pm d/2 \mp t$, where $d$ denotes the distance between the bubble centers. Two important remarks are in order: First, we see that the steep drop after the collision becomes shorter as $\tilde{\lambda}$ decreases. This can be traced to false vacuum trapping (field bouncing back to the false vacuum in the collision region) which becomes increasingly likely for larger values of $\tilde{\lambda}$.\,\footnote{In Ref.~\cite{Lewicki:2020jiv} we used larger values of $\tilde{\lambda}$ and found scaling resembling the left panel of Fig.~\ref{fig:scaling}. Here we focus on $\tilde{\lambda}\ll 1$ which is more relevant for very strong transitions.} Second, the larger $\tilde{g}$ is, the closer the behaviour of the GW source is to the case where the complex phases inside the colliding bubbles are equal, $\Delta\varphi=0$. Moreover, the smaller $\tilde{\lambda}$ is, the faster the scaling reaches the $\Delta\varphi=0$ case as a function of $\tilde{g}$. From Fig.~\ref{fig:P} we see that $\tilde{\lambda}$ is very small, $\tilde{\lambda}\ll 1$, and $\tilde{g}$ is large, $\tilde{g} > 10$, in the region where supercooling is strong and the bubble collision signal can be the dominant contribution. As can be seen from the right panel of Fig.~\ref{fig:scaling} the scaling in this case quickly reaches $\propto t^{-3}$ behaviour after the collision.\footnote{We have also checked that subsequent collisions do not change the scaling.} Instead, for example in the case of breaking of a global U(1) symmetry, corresponding to $\tilde{g}=0$, $\propto t^{-2}$ scaling can be realized.

\section{Gravitational wave spectrum}
Next, following a similar approach as one would using the envelope approximation, we perform many-bubble simulations in the thin-wall limit. Whereas in the envelope approach the collided parts of the walls are neglected, we instead use the scaling obtained in the previous section. 

We consider an exponential bubble nucleation rate per unit volume, $\Gamma \propto e^{\beta t}$, and write the abundance of GWs produced in bubble collisions in a logarithmic frequency interval as~\cite{Lewicki:2020jiv}
\begin{equation} \label{eq:Omega}
\Omega_{\rm GW}(\omega) \equiv \frac{1}{E_{\rm tot}}\frac{\td E}{\td\ln\omega} = \left(\frac{H}{\beta}\right)^2\left(\frac{\alpha}{1+\alpha}\right)^2 S(\omega) \,,
\end{equation}
where 
\begin{equation} \label{eq:S}
S(\omega) \!=\! \left(\frac{\omega}{\beta}\right)^3 \frac{3\beta^5}{8\pi V_s} \int \!\td\Omega_k \left[ |C_+(\omega)|^2 + |C_\times(\omega)|^2 \right]
\end{equation}
gives the spectral shape of the GW background. The volume over which $\Omega_{\rm GW}$ is averaged is denoted by $V_s$. The functions $C_{+,\times}$ are for $\hat{k} = (0,0,1)$, in the thin-wall limit given by (see Appendix~\ref{thinwall} for the derivation) 
\begin{equation}\begin{aligned} \label{eq:Cpc}
C_{+,\times}(\omega) \approx \frac{1}{6\pi} \sum_n &\int_{t_n} \td t\, \td \Omega_x\, \sin^2\theta_x\, g_{+,\times}(\phi_x) \\ &\times R_n^3 f(R_n) \,e^{i\omega (t - z_n - R_n\cos\theta_x)} \,,
\end{aligned}\end{equation}
where $t_n$, $z_n$ and $R_n = t-t_n$ denote the nucleation time, the $z$ coordinate of the bubble nucleation center and the radius of the bubble $n$. The functions $g_{+,\times}$ are defined as $g_+(\phi_x) = \cos(2\phi_x)$ and $g_\times(\phi_x) = \sin(2\phi_x)$. The function $f(R_n)$ accounts for the scaling of the GW source, 
\begin{equation} \label{eq:Edecay2}
f(R_n) = \min\left[1,\left(R_{n,c}/R_n\right)^{\xi+1}\right] \,,
\end{equation}
following the results of our lattice simulations, which showed that the maximum of $T_{zz}$ scales as $R_n^{-\xi}$ after the collision. The bubble radius at the collision moment, $t=t_c$, is denoted by $R_{n,c}$.

\begin{figure}
\centering
\includegraphics[width=0.96\columnwidth]{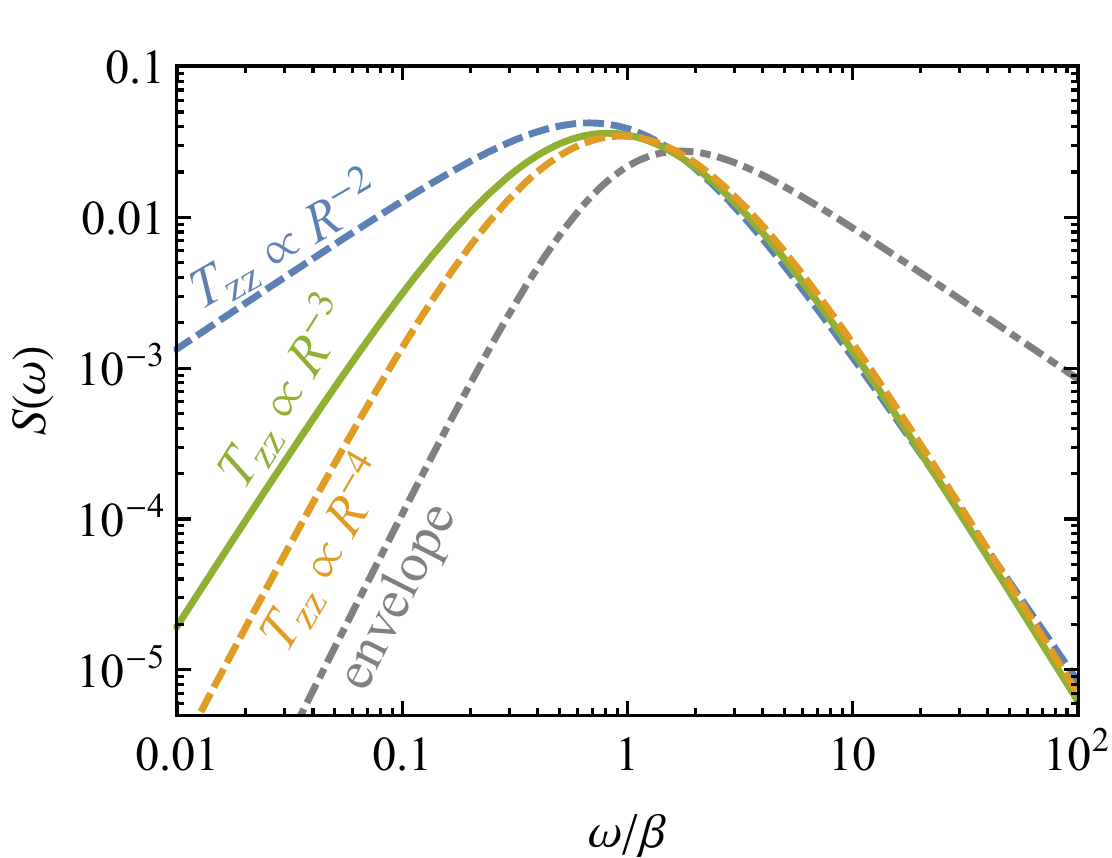}
\caption{The spectral shape of GWs (see Eq.~\eqref{eq:Omega}) from vacuum bubble collisions. The curves show broken power-law fits to the simulation results for different decay-laws of the GW source after collisions and in the envelope approximation. The solid curve is realised in the case of breaking of a gauge U(1) symmetry. The corresponding parameters of the fit, and their errors, are given in Table~\ref{table:fit}.}
\label{fig:S}
\end{figure}

We calculate $S$ by performing thin-wall simulations where we nucleate bubbles according to the rate $\Gamma \propto e^{\beta t}$ a cubic box of size $(7/\beta)^3$ with periodic boundary conditions (see Appendix~\ref{thinwall} for the details of the thin-wall simulations). We perform the angular integrals over the bubble surfaces by discretising each of them with $10^6$ evenly distributed points. Our results are calculated from 60 simulations. We parametrize the results as a broken power-law,
\begin{equation} \label{eq:fit}
S_{\rm fit}(\omega) =\frac{\bar S\,(a+b)^c}{\left[b \left(\frac{\omega}{\bar\omega}\right)^{-a/c} + a \left(\frac{\omega}{\bar\omega}\right)^{b/c}\right]^c} \,,
\end{equation}
where $\bar S$ and $\bar\omega$ are the peak amplitude and angular frequency of the spectrum, $a,b>0$ are the low- and high-frequency slopes of the spectrum respectively, and $c$ determines the width of the peak. We show the parameter values and their errors resulting from fits to our simulation results in Table~\ref{table:fit} for $\xi=2,3,4$ and the envelope approximation,\footnote{The envelope approximation~\cite{Kosowsky:1992vn} corresponds to $f(R_n>R_{n,c})=0$, obtained in the limit $\xi\to\infty$.} and illustrate these fits in  Fig.~\ref{fig:S}. The spectrum today can be obtained from Eq.~\eqref{eq:Omega} by red-shifting~\cite{Kamionkowski:1993fg,Lewicki:2020jiv}. At super-horizon scales the spectrum scales as $\omega^3$ as the source is diluted by the Hubble expansion~\cite{Caprini:2009fx,Cai:2019cdl}.

For $\xi=3$, corresponding to the case of breaking of a gauge U(1) symmetry, the low- and high-frequency tails of the spectrum are $\propto \omega^{2.3}$ and $\propto \omega^{-2.4}$. Instead, for $\xi=2$, which can be realized for example in the case of breaking of a global U(1) symmetry, they are $\propto \omega^{1.0}$ and $\propto \omega^{-2.2}$. The spectrum peaks in both cases slightly below $\omega = \beta$ with an amplitude $S\approx 0.04$. We find that increasing $\xi$ brings the low-frequency power-law quickly closer to envelope result, $a=3.0$, as shown by the $\xi=4$ case. The high-frequency power-law instead seems to change very mildly for $\xi>3$ not obviously converging to a slope that agrees with the envelope approximation.\footnote{We have also calculated the spectrum for $\xi = 5$ in which case $a = 3.02$ and $b=2.39$.}

\begin{table}[h]
\centering
\begin{tabular}{ p{0.78cm} p{1.25cm} p{1.45cm} p{1.45cm} p{1.45cm} p{1.25cm} }
\hline\hline
& $\ \ \, 100 \bar S$ & $\quad \  \bar\omega/\beta$ &$\quad \ \ \, a$ & $\quad \ \ \, b$ & $\quad \ \, c$\\
\hline
$\xi=2$ & $4.23 \pm 0.1$ & $0.68\pm0.01$ & $1.00 \pm 0.02$  & $2.17 \pm 0.05$ & $2.02 \pm 0.1$   \\
$\xi=3$ & $3.61 \pm 0.1$ & $0.82\pm0.01$ & $2.34 \pm 0.03$  & $2.41 \pm 0.02$ & $4.20 \pm 0.2$   \\
$\xi=4$ & $3.46 \pm 0.1$ & $0.93\pm0.01$ & $2.87 \pm 0.04$  & $2.42 \pm 0.02$ & $4.63 \pm 0.2$  \\
env. & $2.75 \pm 0.1$ & $1.72\pm0.04$ & $2.98 \pm 0.02$  & $1.01 \pm 0.02$ & $2.18 \pm 0.1$  \\
\hline\hline
\end{tabular}
\caption{Fitted values for the parametrization of the spectral shape~\eqref{eq:fit}.}
\label{table:fit}
\end{table}

\section{Conclusions} 
Vacuum bubble collisions give the dominant source of GWs in a cosmological first-order phase transition if the transition is sufficiently strongly supercooled. This can be realized in classically conformal models. The simplest realistic examples of such involve breaking of a U(1) gauge symmetry. Motivated by these observations, we have studied the formation of GWs in a first-order phase transition in classically conformal scalar electrodynamics. 

We have estimated the GW spectrum by first studying the scaling of the GW source in two-bubble lattice simulations, and then using that scaling in many-bubble simulations in the thin-wall limit. We have found that the presence of the gauge field brings the results close to the simple real scalar case where the GW source decays with the bubble size as $\propto R^{-3}$. The resulting spectrum, shown by the green solid curve in Fig.~\ref{fig:S}, follows $\Omega_{\rm GW}\propto \omega^{2.3}$ at low frequencies and $\Omega_{\rm GW} \propto \omega^{-2.4}$ at high frequencies. By calculating the transition temperature in classically conformal scalar electrodynamics we have shown that this limit with $\tilde{\lambda} \ll 1$ and $\tilde{g} \gg 1$ is realised in most of the parameter space of interest where the bubble collision signal can give the dominant contribution to the GW spectrum.

Ascertaining the shape of the signal is crucial as it could shine light on the underlying particle physics model. Sources associated to plasma dynamics, that dominate GW production in weaker transitions, in general predict high-frequency slopes~\cite{Caprini:2015zlo} $\omega^{-4}$ from sound waves and $\omega^{-2/3}$ from turbulence. Our result shows that probing a signal that falls between the power-laws $\omega^{-2.5}$ and $\omega^{-2.1}$ at high frequencies would point to a very strong phase transition.

While our result should describe both real and gauged scalar field transitions, we have explored also different energy decay laws which could be realised in other models. Most notably, $T_{zz}\propto R^{-2}$ could be realised in models where the U(1) symmetry is global (i.e. $\tilde{g} = 0$) or modified transition dynamics allows $\tilde{g}\ll 1$. In this case the resulting spectrum follows $\Omega_{\rm GW}\propto \omega$ at low and $\Omega_{\rm GW}\propto \omega^{-2.2}$ at high frequencies. Moreover, we have shown that the envelope result is unlikely to be able to describe realistic spectra especially at high frequencies.

\begin{acknowledgments}
ML was supported by the Polish National Science Center grant 2018/31/D/ST2/02048 and VV by Juan de la Cierva fellowship from Spanish State Research Agency. The project is co-financed by the Polish National Agency for Academic Exchange within Polish Returns Programme under agreement PPN/PPO/2020/1/00013/U/00001. This work was also supported by the grants FPA2017-88915-P and SEV-2016-0588. IFAE is partially funded by the CERCA program of the Generalitat de Catalunya.
\end{acknowledgments}

\onecolumngrid

\appendix

\section{Lattice simulation} \label{lattice}

We perform lattice simulations of two-bubble collisions, starting from a configuration where $A_\mu = 0$ and two O(4) symmetric scalar field bubbles have nucleated simultaneously at $(x,y,z)=(0,0,\pm d/2)$. The radial profile of an O(4) symmetric initial configuration is obtained as the solution of
\begin{equation}
\partial_r^2 |\phi| + \frac3r \partial_r |\phi| = \frac{\td V}{\td |\phi|} 
\end{equation}
with boundary conditions $\partial_r |\phi|=0$ at $r=0$ and $|\phi|\to 0$ at $r\to\infty$. A system of two simultaneously nucleated O(4) symmetric bubbles is conveniently described in coordinates $(s,z,\psi,\theta)$ defined via $\tan\theta = x/y$, $t = s\cosh\psi$ and $r = s\sinh\psi$ where $r^2=x^2+y^2$. We consider the region $t>r$ as this is where the bubbles collide (see Ref.~\cite{Lewicki:2019gmv} for details). The d'Alembertian in these coordinates reads
\begin{equation}
\Box X = \partial_s^2 X + \frac{2}{s} \partial_s X - \partial_z^2 X \,.
\end{equation}
For the lattice implementation, we write the equations of motion of the scalar and gauge fields in dimensionless variables, $\phi' = \phi/v$, $A' = A/v$, $x'^\mu = \sqrt{\Delta V} x^\mu/v$, as
\begin{equation}\begin{aligned} \label{eq:EOM}
&\Box' A'_{s,z} = \tilde{g} \big(\phi'_I \partial'_{s,z} \phi'_R - \phi'_R \partial'_{s,z} \phi'_I\big) - \tilde{g}^2 A'_{s,z} \big({\phi'_R}^2 + {\phi'_I}^2\big) \,, \\
&\Box' \phi'_R + \frac{\td V'}{\td \phi'_R} = 2 \tilde{g}\big( A'_s \partial'_s \phi'_I - A'_z \partial'_z \phi'_I \big) + \tilde{g}^2 \big({A'_s}^2-{A'_z}^2\big) \phi'_R \,, \\ 
&\Box' \phi'_I + \frac{\td V'}{\td \phi'_I} = -2 \tilde{g} \big( A'_s \partial'_s \phi'_R - A'_z \partial'_z \phi'_R \big) + \tilde{g}^2 \big({A'_s}^2-{A'_z}^2\big) \phi'_I \,,
\end{aligned}\end{equation}
where $\phi_R$ and $\phi_I$ are the real and imaginary parts of $\phi$, defined such that $\phi = (\phi_R+i\phi_I)/\sqrt{2}$. The dimensionless scalar potential $V' = V/\Delta V$, where $\Delta V$ denotes the vacuum energy difference between the symmetric and broken vacua at temperature $T$, can be written as
\begin{equation}
V'(\phi') = \tilde{\lambda} |\phi'|^2 \!+\! |\phi'|^4 \left[(\tilde{\lambda}\!+\!2)\ln|\phi'|^2 \!-\! (\tilde{\lambda}\!+\!1)\right] \,.
\end{equation}
Here we have defined dimensionless parameters $\tilde{g}$ and $\tilde{\lambda}$ as
\begin{equation} \label{eq:kappaG}
\tilde{g} = \frac{g v^2}{\sqrt{\Delta V}} \,, \qquad \tilde{\lambda} = \frac{g^2 v^2 T^2}{2 \Delta V} \,.
\end{equation}
We solve the equations of motion~\eqref{eq:EOM} numerically on a diamond-shaped $sz$ lattice as in Ref.~\cite{Hawking:1982ga}. To ascertain the numerical stability of the simulation, we have checked that the gauge condition, $\partial^\mu A_\mu = 0$ remains satisfied througout the simulation. We have also performed the simulations with different lattice spacings finding that the results are unchanged unless the grid is much less dense than what we use in the following results ($\delta s' = \delta z' = 0.005$).

\begin{figure*}
\centering
\includegraphics[width=0.98\textwidth]{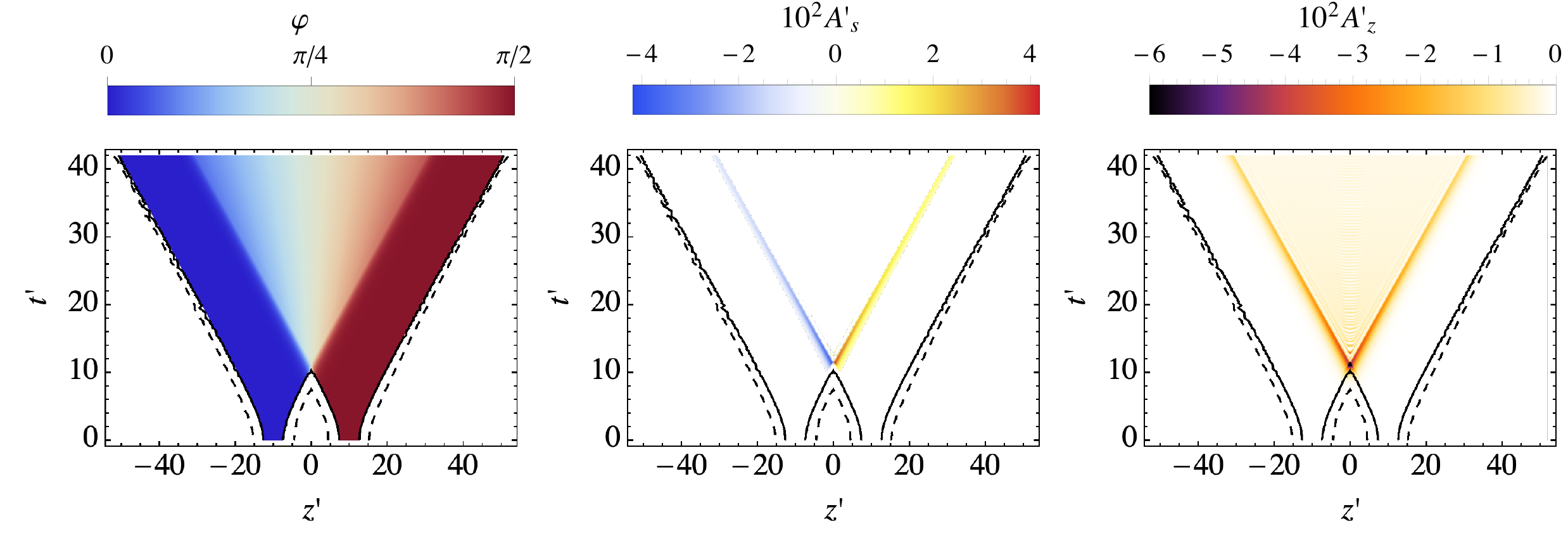}
\caption{Collision of two bubbles with initial complex phase difference $\Delta\varphi=\pi/2$. The solid and dashed curves correspond to $|\phi'|=0.1$ and $|\phi'|=0.01$, respectively. The color coding indicates in the left panel the complex phase of the scalar field, in the middle panel the $s$ component of the gauge field and in the right panel the $z$ component of the gauge field.}
\label{fig:phiA}
\end{figure*}

In Fig.~\ref{fig:phiA} we show the result from a simulation with $\tilde{\lambda}=0.04$, $\tilde{g}=10$, initial bubble separation $d'=20$ (in the dimensionless units) and initial complex phase difference $\Delta\varphi=\pi/2$. The left panel shows the evolution of the complex phase of the scalar field. As can be seen from the equations of motion~\eqref{eq:EOM}, gradients in $\varphi$ source the gauge field. Therefore, it is expected that the gauge field deviates from zero where the gradients in $\varphi$ are large. We see this in the middle and right panels of Fig.~\ref{fig:phiA}, which show the gauge field components $A_s$ and $A_z$: A sharp feature in the gauge field propagates roughly at the speed of light after collision.

\begin{figure*}
\centering
\includegraphics[width=0.94\textwidth]{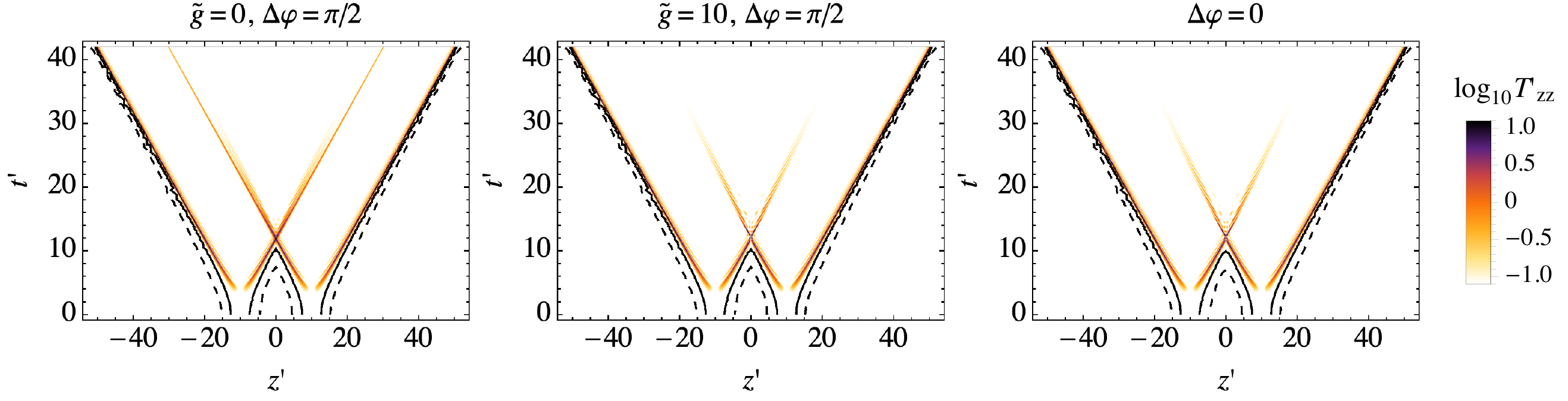}
\caption{Collision of two bubbles with values of $\Delta\varphi$ and $\tilde{g}$ indicated above the plots, and  $\tilde{\lambda}=0.04$. For the $\Delta\varphi=0$ case the value of $\tilde{g}$ is irrelevant. The solid and dashed curves correspond to $|\phi'|=0.1$ and $|\phi'|=0.01$, respectively, and the color coding indicates the value of the $zz$ component of the stress energy tensor.}
\label{fig:Tzz}
\end{figure*}

In Fig.~\ref{fig:Tzz} we show the $zz$ component of the stress-energy tensor,
\begin{equation}
T'_{zz} = \,(\partial'_z{\phi'_R})^2+(\partial'_z{\phi'_I})^2 - \partial'_z A'_s\left(\partial'_z A'_s - \partial'_s A'_z\right) + \tilde{g} A'_z (\phi'_I \partial'_z \phi'_R - \phi'_R \partial'_z \phi'_I) \,,
\end{equation}
by the color coding for three different two-bubble collisions. In the left panel $\tilde{g} = 0$, and in the right panel the complex phase of the scalar field inside the colliding bubbles is the same, $\Delta\varphi = 0$. In these cases only the scalar field gradients contribute to $T'_{zz}$, and the result agrees with the ones shown in Ref.~\cite{Lewicki:2020jiv}: If there is a complex phase difference between the colliding bubbles, the scalar field gradients propagate much longer after the collision than in the case where the complex phases are equal. The middle panel of Fig.~\ref{fig:Tzz} shows the case where the complex phases are different and $\tilde{g}>0$. We see that the result in that case roughly matches with the $\Delta\varphi = 0$ case. This can be understood as decay of the gradients in the complex phase of the scalar field to gauge fields.

The second crucial piece of information we get from Fig.~\ref{fig:Tzz} is that the gradients are well localised in space not only as the walls accelerate and become thinner but also after the time of the collision. In fact the spatial localisation of gradients becomes even more narrow as bubbles grow bigger before colliding. In a realistic transition the bubbles would grow many orders of magnitude in size before colliding which means it is well justified to assume the thickness of the walls and gradients after the collision is negligible compared to size of the colliding bubbles. This is the well known thin-wall approximation we will utilise in Appendix~\ref{thinwall}.

\section{Thin-wall simulation} \label{thinwall}

Next we generalize the treatment of Ref.~\cite{Lewicki:2020jiv} to the case where the stress-energy tensor is not given solely by the scalar field gradients after the bubble collisions. The Fourier transform of the stress-energy tensor is given by
\begin{equation}
T_{ij}(\vec{k}) = \frac{1}{2\pi} \int\td t \,\td^3x\, e^{i\omega t-i\vec{k}\cdot\vec{x}} \,T_{ij}(\vec{x}) \,.
\end{equation}
In the thin wall limit, by breaking the spatial integral into regions around each bubble nucleation center and taking $\vec{k} = (0,0,\omega)$, we get
\begin{equation}
T_{ij}(\vec{k}) = \frac{1}{2\pi} \sum_n \int_{t_n}\!\td t \,\td\Omega_x e^{i\omega (t - z_n - R_n\cos\theta_x)} \!\int\!\td r\, r^2 T_{ij}(r) \,,
\end{equation}
where $t_n$ denotes the nucleation time of the bubble $n$, $z_n$ the $z$ coordinate of the bubble nucleation center, and $R_n=t-t_n$ the bubble radius at time $t>t_n$. It is convenient to write $T_{ij}(r)$ in a coordinate system where the $z$ axis points to the radial direction, $\hat{z} = \hat{r}$. The coordinate transformation of $T_{ij}(r)$ is
\begin{equation}
T_{ij}(r) = T_{i'j'}(r) \frac{\partial x_{i'}}{\partial x_i} \frac{\partial x_{j'}}{\partial x_j} \,.
\end{equation}
If only $zz$ component of is non zero, we get
\begin{equation}
\int \!\td r\, r^2 \,T_{ij}(r) = \hat x_i \hat x_j \int \td r\, r^2 T_{zz}(r) \,.
\end{equation}
In the thin-wall limit we approximate
\begin{equation}
\int\! \td r\, r^2 T_{zz}(r) \approx L R_n^2 \max\!\left[T_{zz}(R_n)\right] \equiv \frac{\Delta V}{3} \!R_n^3 f(R_n) \,,
\end{equation}
where $L$ denotes the bubble wall width. 

Before the wall element in the solid angle $\td \Omega$ collides with another bubble $T_{zz}$ is given solely by the scalar field gradients, $T_{zz} = |\partial_z \phi|^2$. By energy conservation
\begin{equation}
\int \! \td r\, r^2 \,|\partial_z \phi|^2 = \frac{\Delta V}{3} R_n^3 \,,
\end{equation}
where $R_{n,c}$ denotes the bubble radius at the time of the collision. Therefore, before the collision 
\begin{equation}
f(R_n\leq R_{n,c})=1 \,.
\end{equation}
From the definition of $f$ and using $f(R_n=R_{n,c})=1$ we get that the wall width at the collision moment is
\begin{equation}
L_c = \frac{\Delta V}{3\max\!\left[T_{zz}(R_{n,c})\right]} R_{n,c} \,.
\end{equation}
Before the collision the bubble wall gets thinner as the Lorentz factor of the bubble wall increases, but after the collision the wall element moves roughly at a constant velocity as no more energy is injected into it. Therefore we assume that after the collision the wall thickness $L$ remains constant, $L=L_c$. We can then write the $f(R_n)$ function after the collision as
\begin{equation}
f(R_n>R_{n,c}) = \frac{3 L_c \max\!\left[T_{zz}(R_n)\right]}{R_n \Delta V} = \frac{R_{n,c}}{R_n} \frac{\max\!\left[T_{zz}(R_n)\right]}{\max\!\left[T_{zz}(R_{n,c})\right]} \,.
\end{equation}
Since $T_{ij}(\vec k)$ is symmetric, we can write the transverse-traceless projection for $\hat{k} = (0,0,1)$ as
\begin{equation} \label{eq:TTproj}
2\Lambda_{ij,lm} T_{ij}^*(\vec{k}) T_{lm}(\vec{k}) \!=\! \Delta V^2 \left(|C_+(\omega)|^2 + |C_\times(\omega)|^2\right) .
\end{equation}
The functions $C_+ \equiv T_{11} - T_{22}$ and $C_\times \equiv 2T_{12}$ are given by
\begin{equation}
C_{+,\times}(\omega) \approx \frac{1}{6\pi} \sum_n \int_{t_n} \td t\, \td \Omega_x\, \sin^2\theta_x\, g_{+,\times}(\phi_x) R_n^3 f(R_n) \,e^{i\omega (t - z_n - R_n\cos\theta_x)} ,
\end{equation}
where $g_+(\phi_x) = \cos(2\phi_x)$ and $g_\times(\phi_x) = \sin(2\phi_x)$. 

The total energy spectrum in a direction $\hat{k}$ at an angular frequency $\omega = |\vec{k}|$ of the GWs emitted in the phase transition is given by~\cite{Weinberg:1972kfs}
\begin{equation}
\frac{\td E}{\td\Omega_k\td \omega} = 2G\omega^2 \Lambda_{ij,lm}(\hat{k}) T_{ij}^*(\vec{k}) T_{lm}(\vec{k}) \,.
\end{equation}
Using Eq.~\eqref{eq:TTproj} and the definition $\alpha\equiv \Delta V/\rho_{\rm rad}$ we write the abundance of GWs produced in bubble collisions in a logarithmic frequency interval as
\begin{equation}
\Omega_{\rm GW}(\omega) \equiv \frac{1}{E_{\rm tot}}\frac{\td E}{\td\ln\omega} = \left(\frac{H}{\beta}\right)^2\left(\frac{\alpha}{1+\alpha}\right)^2 S(\omega) \,,
\end{equation}
where $E_{\rm tot} = V_s(\rho_{\rm rad} + \Delta V)$ and
\begin{equation}
S(\omega) = \left(\frac{\omega}{\beta}\right)^3 \frac{3\beta^5}{8\pi V_s} \int \!\td\Omega_k \left[ |C_+(\omega)|^2 + |C_\times(\omega)|^2 \right]
\end{equation}
gives the spectral shape of the GW background. Here $V_s$ denotes the volume over which $\Omega_{\rm GW}(\omega)$ is averaged. We consider exponential bubble nucleation rate, $\Gamma \propto e^{\beta t}$, which implies that $\int \!\td\Omega_k [ |C_+(\omega)|^2 + |C_\times(\omega)|^2 ] \propto V_s/\beta^5$.

We simulate the phase transition by nucleating bubbles according to the exponential bubble nucleation rate inside a cubic box with periodic boundary conditions. Following the thin-wall approximation, we simulate the bubbles as spherical shells. We discretise the bubble surfaces and find the time when each of these bubble wall elements collides with another bubble wall by bisection method. The corresponding radius is denoted by $R_n = R_{n,c}$. Once we know $R_{n,c}$ for each bubble wall element of each bubble in the simulation, we integrate the functions $C_{+,\times}(\omega)$ for a given value of $\omega$. We note that if $f(R)$ is a (broken) power-law the temporal integral can be performed analytically. The spectral function $S(\omega)$ is then simply obtained by integrating over the $\hat{k}$ directions. In practice, since our simulation box is cubic, the integral over $\hat{k}$ directions is done by summing over 6 directions, corresponding to the normal vectors of the faces of the cube, with equal weights $2\pi/3$. Finally, to reduce the errors, we calculate the final result by averaging the spectrum over many simulations with different randomly generated bubble nucleation histories.

\bibliography{gw}

\end{document}